\documentclass{PoS}
\usepackage{amsmath} 
\usepackage{lineno}

\title{Monte Carlo Studies of the GCT Telescope for the Cherenkov Telescope Array}

\ShortTitle{MC Simulations for GCT}

\author{\speaker{Thomas Armstrong}\\
        Dept. of Physics and Centre for Advanced Instrumentation, Durham University\\
        E-mail: \email{thomas.armstrong@durham.ac.uk}}

\author{Heide Costantini\\
        Aix Marseille Universit\'{e}, CNRS/IN2P3, CPPM UMR 7346, 13288 Marseille, France\\
        E-mail: \email{costant@cppm.in2p3.fr}}

\author{Cameron Rulten\\
               LUTH, Observatoire de Paris, CNRS, Universit\'{e} Paris Diderot, PSL, France \\
		now at: School of Physics and Astronomy, University of Minnesota, 116 Church St SE, Minneapolis, MN 55455, USA\\
		E-mail: \email{rulten@physics.umn.edu }}

\author{Victor Stamatescu\\
        University of Adelaide, SA 5005, Australia\\
        E-mail: \email{victor.stamatescu@gmail.com}}

\author{Andreas Zech\\
        LUTH, Observatoire de Paris, CNRS, Universit\'{e} Paris Diderot, PSL, France\\
        E-mail: \email{andreas.zech@obspm.fr }}
        
\author{For the CTA Consortium\\}

\abstract{The GCT is an innovative dual-mirror solution proposed for the small-size telescopes for CTA, capable of imaging primary cosmic gamma-rays from below a TeV to hundreds of TeV. The reduced plate scale resulting from the secondary optics allows the use of compact photosensors, including multi-anode photomultiplier tubes or silicon photomultipliers. We show preliminary results of Monte Carlo simulations using the packages \texttt{CORSIKA} and \texttt{Sim\_telarray}, comparing the relative performance of each photosensor type.  We also investigate the effect of the secondary optics in terms of optical performance, image resolution and camera response.  With the ongoing commissioning of the  prototype structure and camera, we present the preliminary expected performance of GCT.}

\FullConference{The 34th International Cosmic Ray Conference,\\
		30 July- 6 August, 2015\\
		The Hague, The Netherlands}

\begin{document}
%\linenumbers
\section{Introduction}

The Cherenkov Telescope Array (CTA)~\cite{cta} will be the next generation observatory for very high energy (VHE) gamma ray astronomy. Running largely as an open observatory, CTA will provide the community with a wealth of data on some of the most extreme objects in the Universe at sensitivities at least an order of magnitude greater than that of current Imaging Atmospheric Cherenkov Telescopes (IACTs, e.g. H.E.S.S, MAGIC and VERITAS) over an energy range of a few tens of GeV to hundreds of TeV. CTA will consist of an observatory in each hemisphere. The southern array, capable of observing towards the centre of the Galaxy, will be the larger of the two, comprising $\sim$100 telescopes of three different sizes in order to cover the required energy range. The showers produced by the most energetic gamma rays and cosmic rays above $\sim$10's of TeV are rare; however, the light yield from these cascades is large enough to allow telescopes to detect them at distances greater than the typical Cherenkov light pool. This allows us to cover a larger area, therefore increasing our detection rates, by building many small-size telescopes (SST). 

There are currently three prototype designs for the SST~\cite{sst}. Here we discuss the Gamma-ray Cherenkov Telescope (GCT). Unlike current IACTs, GCT will be based on the Schwarzschild-Couder design~\cite{Vas2007}, a two mirror concept which provides many benefits. The small focal length allows us to obtain a large angular acceptance per pixel using relatively small pixels, while the two mirror system enables us to achieve a PSF within that scale to field angles up to 4.5$^{\circ}$. All together this means that we can reach a Field of View of 9$^{\circ}$ using commercially available compact photo-sensors of size 6$\times$6~mm$^{2}$.

\section{GCT: Optical Structure and Camera}
\label{sec:gct}

%\subsubsection*{Optical structure}

%The mechanical structure of the GCT includes a tower supporting the altitude-azimuth drive system, a dish holding the primary mirror (M1), and a mast and truss 
%structure to carry the secondary mirror (M2). The mast and truss structure consists of eight arms made of thin steel tubes
%in a Serrurier configuration and four braces that provide reinforcement in the vertical direction. The camera is attached to the M2 support with 
%two main arms and two thinner tubes that provide additional stiffness. 
%
%

The optical structure of GCT includes a dish holding the primary mirror (M1), and a mast and truss structure to carry the secondary mirror (M2). The mast and truss structure consists of eight arms made of thin steel tubes and four vertical braces. The camera is attached to the M2 support with two main arms and two thinner tubes that provide additional stiffness. For a full description of the structure see \cite{structure}. The surfaces of the two mirrors are described by 16th order polynomials, resulting in an optimal PSF over the whole field of view. M1, with a diameter of 4\,m, is made up of six petals. A central hole of 130\,cm diameter, which lies in the shadow of the M2 for all field angles, allows the integration of a laser system for telescope alignment. 
After correction for shadowing by the M2, the reflecting area of M1 is about 9.2 m$^2$. M2 will be formed as a monolithic mirror, with a diameter of 2\,m and reflecting area of 3.3 m$^2$. It will contain a small central hole with a diameter of 10\,cm that will be equipped with hardware for alignment, pointing and calibration.

Due to the availability of commercial photo sensors, GCT is currently prototyping two versions of its camera~\cite{chec2}. One using Multi-Anode Photo-Multipliers Tubes (MAPMTs) and the other Silicon Photo-Multipliers (SiPM), from here on referred to as GCT-M and GCT-S respectively. The cameras is comprised of 32 modules, each of these consists of: The photo sensor modules; a pre-amplifier which amplifies and shapes the signals; readout electronics based on the TARGET~\cite{target} module which samples and digitizes the incoming waveform. These 32 modules then feed into a backplane that organises the trigger and then finally sends the data to the camera server via the DACQ.

Throughout this whole chain, from mirrors to the DACQ, there are many variables that we must consider alongside the effect that these can have on the final performance of CTA. 
Therefore, in these proceedings, we present initial detailed Monte Carlo (MC) simulations for GCT with the purpose of estimating the optical performance in terms of shadowing, equivalent area and the point spread function in Section \ref{sec:opt}. Furthermore we consider the charge resolution in Section \ref{sec:cr} and trigger efficiency in Section \ref{sec:trigeff}.
 
\section{Monte Carlo Simulation Chain}
\label{intro:MC}

In order to arrive at an understanding of the system, we need to simulate the entire process, starting from the initial interaction of the primary gamma ray or cosmic ray with the atmosphere, to the arrival of the photons at the focal plane of the camera and then the electronics of the camera itself. This achieved with the two programs \texttt{CORSIKA} and \texttt{sim\_telarray}~\cite{corsika,Ber2008}.

{\bf \texttt{CORSIKA}}: The simulation software \texttt{CORSIKA} is an open source cosmic ray propagation tool. Using detailed electromagnetic and hadronic models, the cascade of secondary particles developing in the atmosphere are tracked. With the "IACT/ATMO" package, we are then able to define an array of telescope positions and sizes, represented by spheres. When the resulting Cherenkov light arrives at ground level, only photons that are incident with these spheres are recorded. 

{\bf \texttt{Sim\_telarray}}: The simulation package \texttt{sim\_telarray} reads in the information produced by \texttt{CORSIKA} and traces the photons from their point on the sphere through the optical assembly to the camera focal plane. Taking into account the shadowing of the structure, the reflectivity of the mirrors and the angular acceptance of detectors or any windows,  
the surviving photons are then converted to electronic signal depending on the input photon detection efficiency (PDE) and single photo-electron (SPE) response, triggered, digitised and read out. Besides the \texttt{CORSIKA} data, the main inputs for \texttt{sim\_telarray} are the configuration files that represent the optics and camera.

\begin{figure}
\centering
\centerline{
\includegraphics[width=1.2\linewidth]{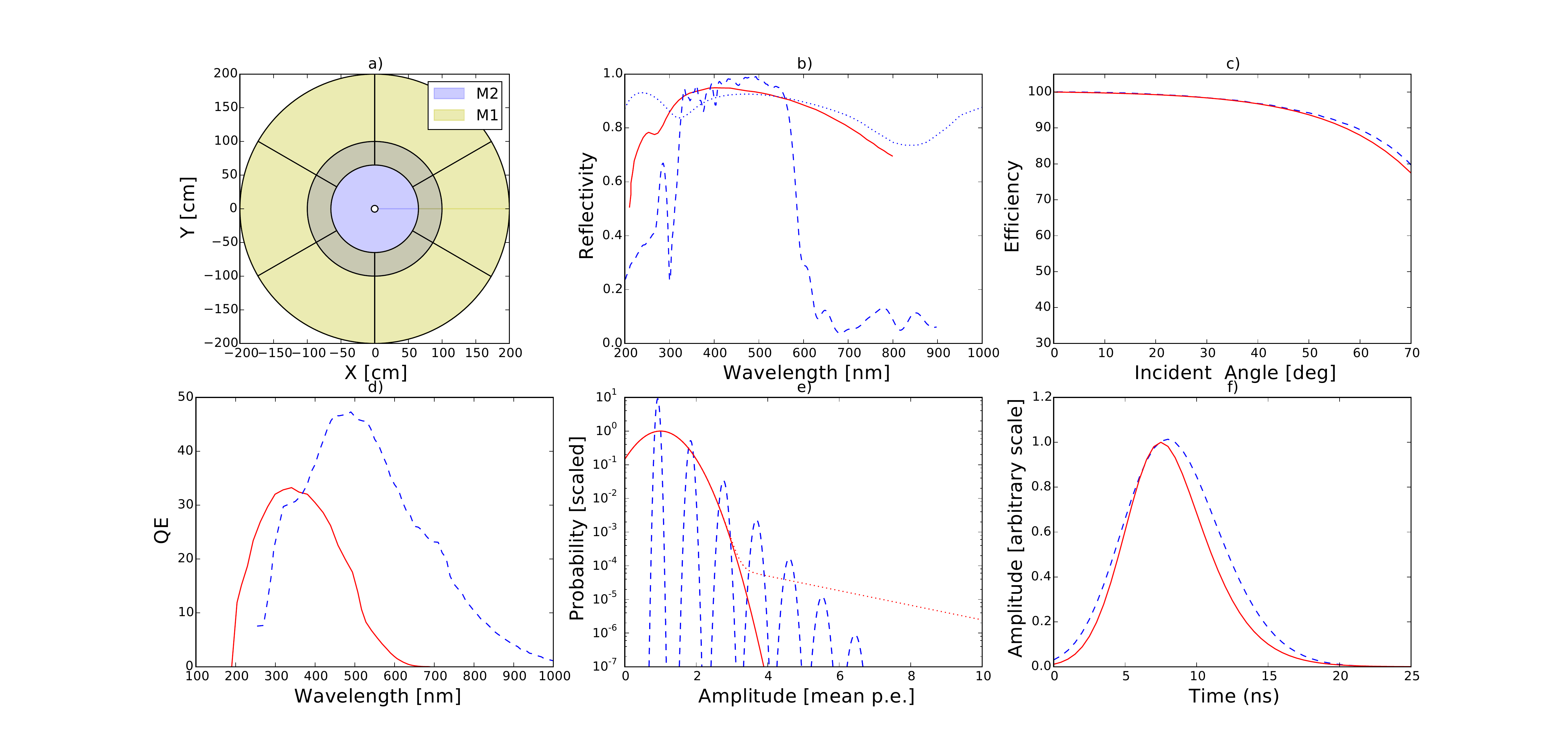}
}
\caption{\small{The following figures red (solid) and blue (dashed) indicates GCT-M and GCT-S receptively a) Geometry of the two mirrors M1 (yellow) and M2 (blue). b) Mirror reflectivity, GCT-M will use an AlSiO coating while GCT-S will use an additional dielectric coating for the primary mirrors (dashed) as well as AlSiO coating for the secondary (dots). c) The angular acceptance. d) The photon detection efficiency of the two detectors. e) Single Photo-Electron response where the dotted line indicates the effect of after pulsing. f) The input to the first level trigger and the sampling rate pulse shape.}}
\label{fig:conf}
\end{figure}

{\bf Telescope Configuration}: Concerning the optical part of the configuration files, there are several main parameters. The two mirror sets, M1 and M2 (Fig. \ref{fig:conf} a), each have their own defined reflectivity as a function of wavelength. GCT-M will use AlSiO coated mirrors for both M1 and M2 while GCT-S is currently set to use a AlSiO coating for M2 and a dielectric coating on M1 (Fig \ref{fig:conf} b). Photons arriving at large incidence angles are then rejected according to the angular acceptance, for GCT-M this concerns the detector itself while GCT-S is mostly affected by a protective window\footnote{There is also a corresponding wavelength dependence of the window, however it is flat over the wavelengths of interest and is not shown} (Fig \ref{fig:conf} c). A certain number of photons do not reach the focal plane as they are blocked by the secondary mirror, camera housing, trusses and masts. This will be discussed further in Section \ref{sec:opt}.

Once the Cherenkov photons enter the detector, they are converted into an electrical signal based on two main parameters. Firstly the PDE, which is related to the cathode currents produced as a function of the incident photons wavelength, and secondly the SPE response which deals with the fluctuation of the amplitude at the anode (see Figures \ref{fig:conf} d and \ref{fig:conf} e). While GCT-S has a larger total QE and improved SPE, its peak sensitivity occurs at longer wavelengths. This results in a larger Night Sky Background (NSB) rate which necessitates the use of dielectric mirrors or a similar coating on the camera window to provide a cut-off at around 550~nm.  The input into the first level trigger, based on the amplification and shaping of the preamp, can be seen in Figure \ref{fig:conf} f. The trigger is then based on this, a defined threshold value and the number of pixels required in the trigger. The same pulse shape is also used for the digitization of the signal, assuming a TARGET ASIC sampling rate of 1~GHz and a readout window of 96 ns (adjustable in 32 ns steps). Due to some similarities in the design of GCT-S and the second two mirror SST concept, ASTRI~\cite{sst}, several parameters here are common between them (namely, mirror reflectivity, window transmittance ($\theta,\lambda$), SPE and PDE) and were provided by ASTRI.

\section{Optical Simulations}
\label{sec:opt}

%\subsection{Shadowing}
%\label{sec:shade}
 
{\bf Shadowing:} When evaluating the effect of shadowing, the main obscuring element to consider is the back of the secondary mirror. Given the design of 
the GCT, the central 25\% of the photon flux within the diameter of M1 are blocked by the back of M2. The reflecting surface of M1 is ring-shaped to 
accommodate this effect. In addition, a smaller fraction of photons are blocked by different elements of the telescope structure in the light path, i.e.\ the camera body,
a protective windshield on the camera body that holds the opened camera lid during operation, and the masts and trusses supporting the M2 structure and camera. 

\begin{figure}
\begin{tabular}{cc}
\includegraphics[width=0.5\textwidth]{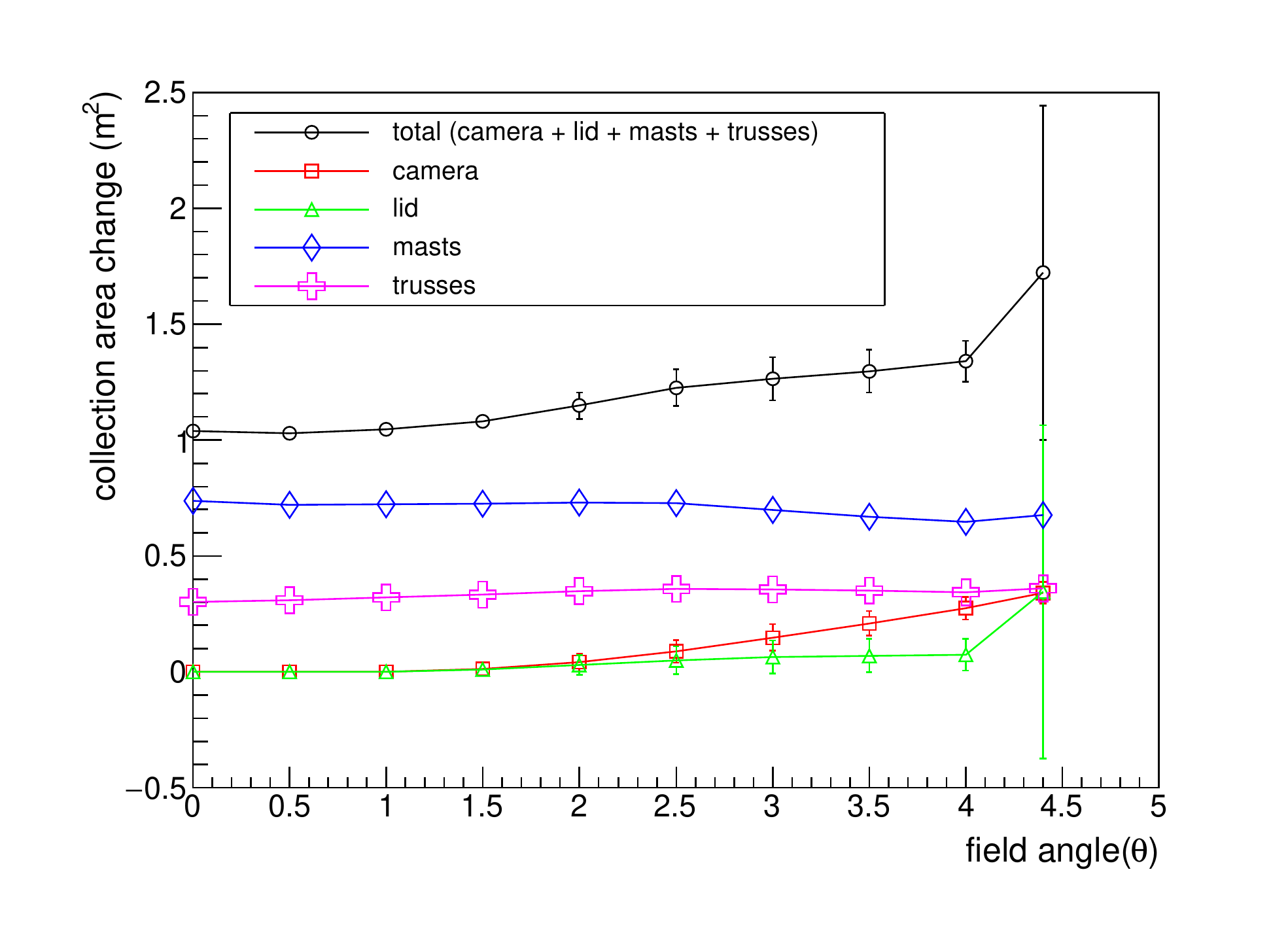} & \includegraphics[width=0.525\textwidth]{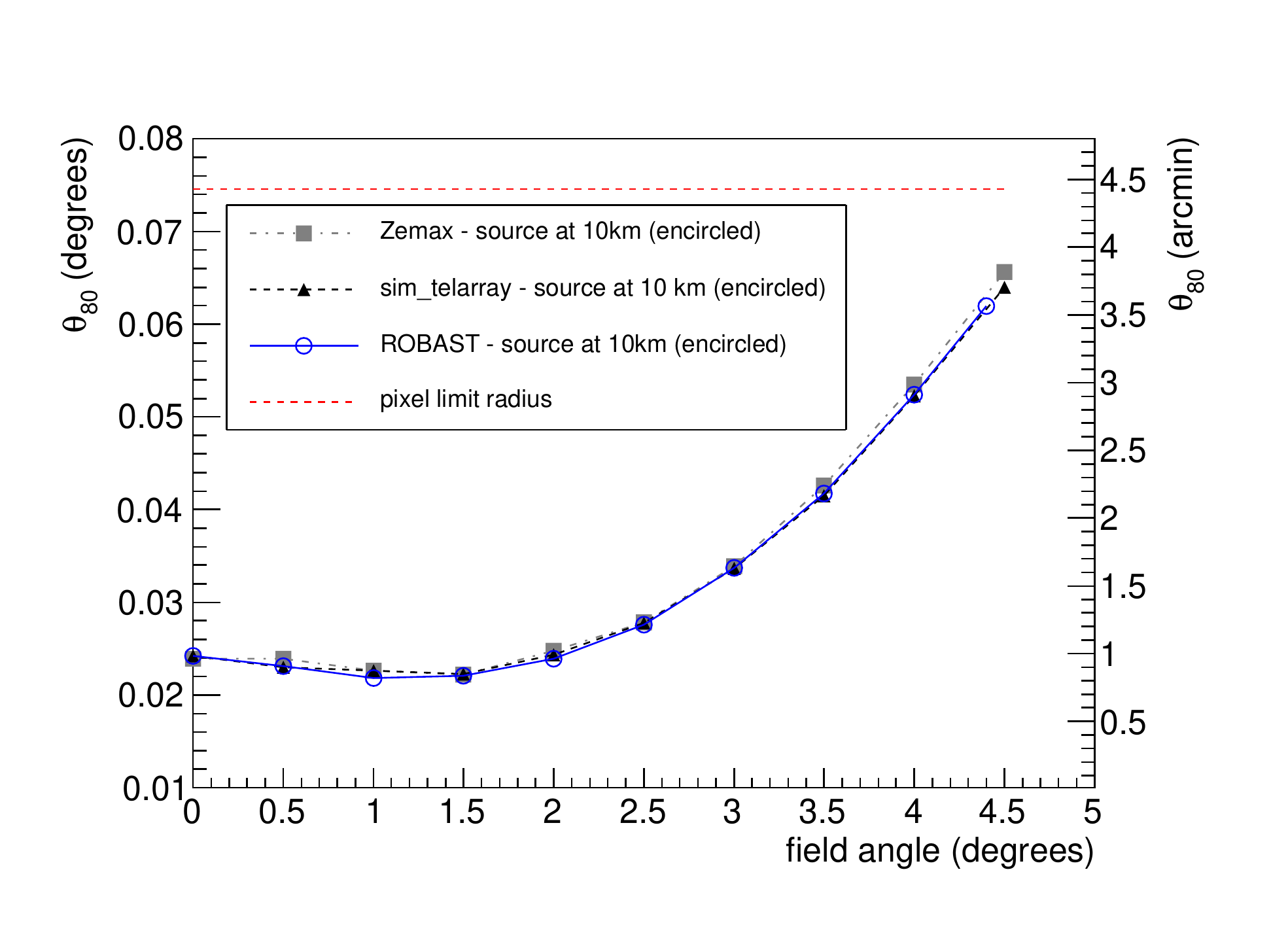}\\
%a) & b) \\
\end{tabular}
  \caption{\small{Left: Change in the collection area due to shadowing of the structural components vs. field angle. Ray-tracing was done using the \texttt{ROBAST} software package~\cite{Oku2011}. Each point is averaged over eight azimuth angles, which explains the large spread in the last point, where the asymmetric effect from the camera lid can be seen. Right: Simulated size of the 80\% containment radius of the PSF as a function of the field angle for a point source at 10\,km. Results for three ray-tracing packages are shown here. For more details see~\cite{Rul2015}.}}
\label{fig:opt}
\end{figure}

The combined shadowing effect of these structural elements and their separate contributions are shown in Fig.~\ref{fig:opt}. For small field angles, the only 
significant contribution comes from the masts and trusses, while the camera housing and the opened camera lid become non-negligible for light sources closer to the edge 
of the field of view. The overall effect is a reduction in the collection area of about 1\,m$^2$ on-axis, increasing to about 1.3\,m$^2$ at 4$^{\circ}$. This corresponds to a shadowing fraction of $\sim$11\% on-axis increasing to $\sim$14\% at 4$^{\circ}$ off-axis.

%
%\subsection{Optical Efficiency / Equivalent area}
%\label{sec:ea}

\begin{figure}[h]
  \centering
  \begin{tabular}{cc}
  \includegraphics[width=0.5\columnwidth]{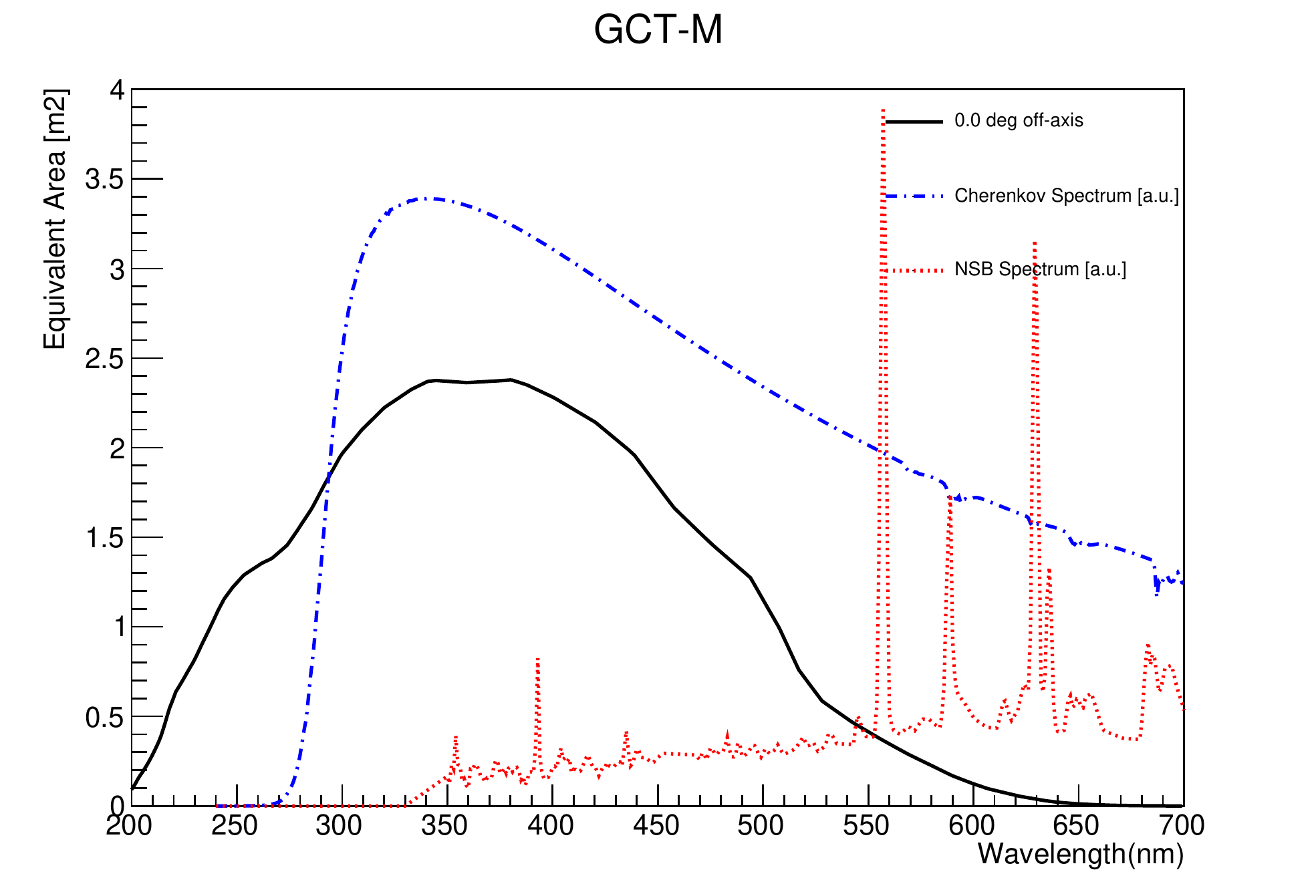} & \includegraphics[width=0.5\columnwidth]{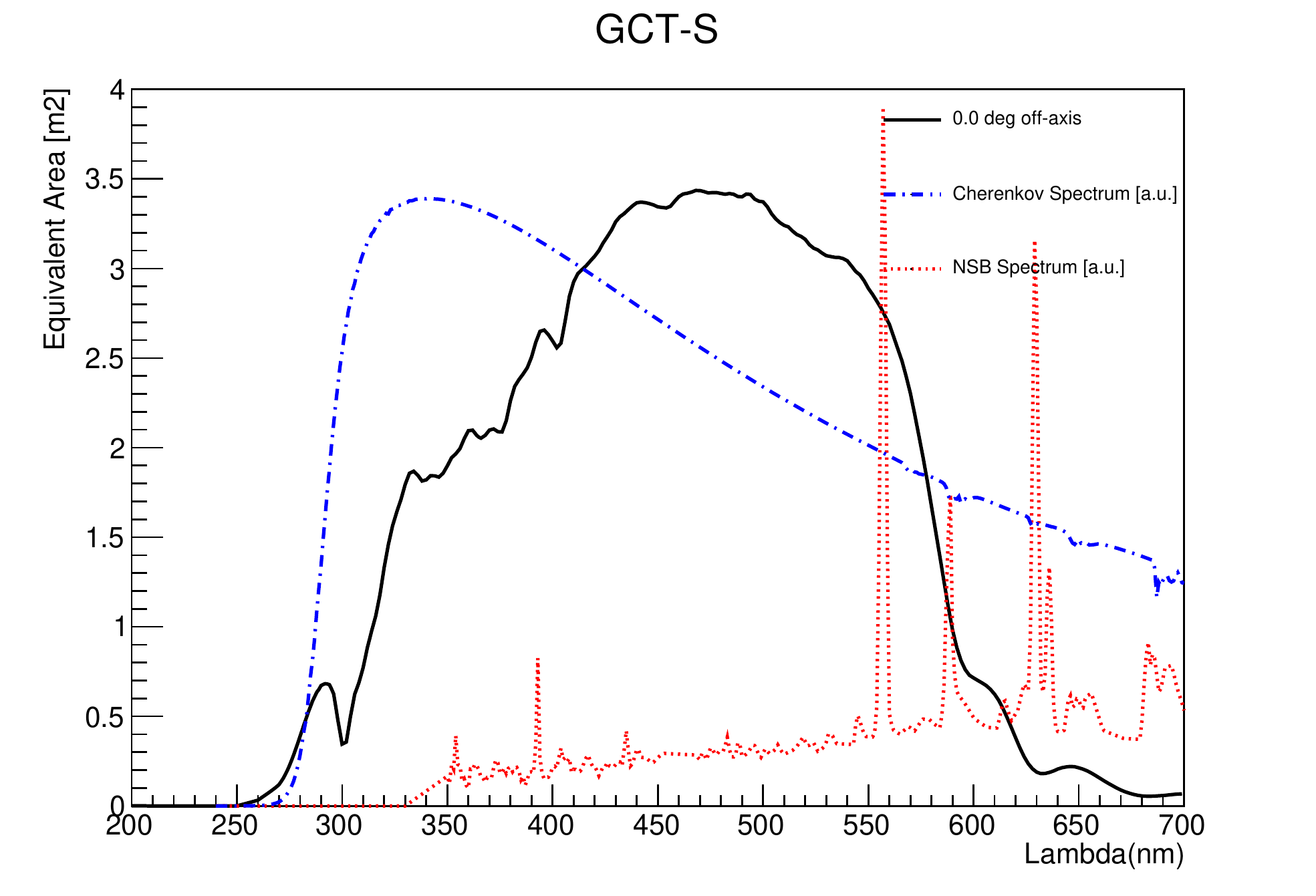} \\
%a) & b) \\
  \end{tabular}
  \caption{\small{Equivalent area of GCT-M (Left) and GCT-S (Right) as a function of the photon wavelength for 0.0$^{\circ}$ off-axis angle. The NSB and Cherenkov spectrum (red and blue dashed line) in arbitrary units have been placed in these figures to show the relative sensitivity ranges of GCT-M and GCT-S. }}
  \label{fig:effective-area-GCTM}
\end{figure}

{\bf Optical Efficiency / Equivalent area:} The equivalent area can be defined as the ratio between the photons detected by the pixel sensors and the photons directed to hit M1 multiplied by the area of M1. 
This parameter accounts for the shadowing of the different telescope components, for the reflectivities of both mirrors (see Fig \ref{fig:conf} b) the transmission of any protective window on the camera, the angular efficiency of the window and sensors (see Fig \ref{fig:conf} c) and finally for the photon detector efficiency of the sensors (see Fig \ref{fig:conf} d).  
The equivalent area has been calculated for both telescopes GCT-M and GCT-S (See Fig \ref{fig:effective-area-GCTM}).
 
%\subsection{Point Spread Function}
% \label{sec:psf}
 {\bf Point Spread Function:} An essential characteristic of the optical performance is the point spread function (PSF) for on-axis light sources and its variation with field angle. We have simulated the ideal PSF with different ray-tracing programs --- \texttt{ZEMAX}, \texttt{ROBAST} and \texttt{sim\_telarray}~\cite{Ber2008} --- and found the results to be in excellent agreement with each other. These simulations were made for a point source at a distance of 10\,km, corresponding to the approximate distance of vertical TeV air showers. 
 A measure to characterize the PSF is the 80\% containment radius $\theta_{80}$, which is shown in Fig.~\ref{fig:opt} as a function of field angle. It can be seen that 
 the PSF is very compact for small angles, with an ideal $\theta_{80}$ of about 0.02$^{\circ}$, compared to an assumed pixel half-width of 0.075$^{\circ}$. The size of the ideal $\theta_{80}$ remains well below the pixel size up to the edge of the field of view.

\section{Charge Resolution}
\label{sec:cr}
The measurement of the time-integrated Cherenkov intensity is a critical task for the camera pixel hardware and/or off-line analysis. The total charge collected by the detector can be expressed as a number of photo-electrons (p.e.) and should be proportional to the Cherenkov light intensity in the pixel. 
The charge resolution is defined as the standard deviation of the reconstructed charge ($\sigma_Q$) around the simulated true charge (Q) and takes into account both the spread of the reconstructed charge distribution and the bias in respect of the simulated true charge.
The acquisition system~\cite{target} allows storage of 96 ns trace for triggered showers in each camera pixel. As the pulse FWHM is much smaller than 96 ns (see Figure \ref{fig:conf} f) to reduce the effect of the NSB photons and electronic noise, it is desirable to integrate over a shorter window around the maximum of the trace to obtain the charge of the incident photon. In the case when the Cherenkov photons and the NSB photons have similar amplitudes,  the NSB contribution can prevent the correct determination of the trace maximum if one looks for the maximum of the signal in the 96 ns window. Therefore the time of the trace maximum has been predicted  from the time of the maximum of the weighted sum of the signals in the neighbour pixels and the specific pixel. Finally the charge resolution has been obtained  for both GCT-S and GCT-M telescopes using a 12 ns integration window around the predicted time of the trace maximum.

\begin{figure}
\begin{tabular}{cc}

\includegraphics[width=0.5\textwidth]{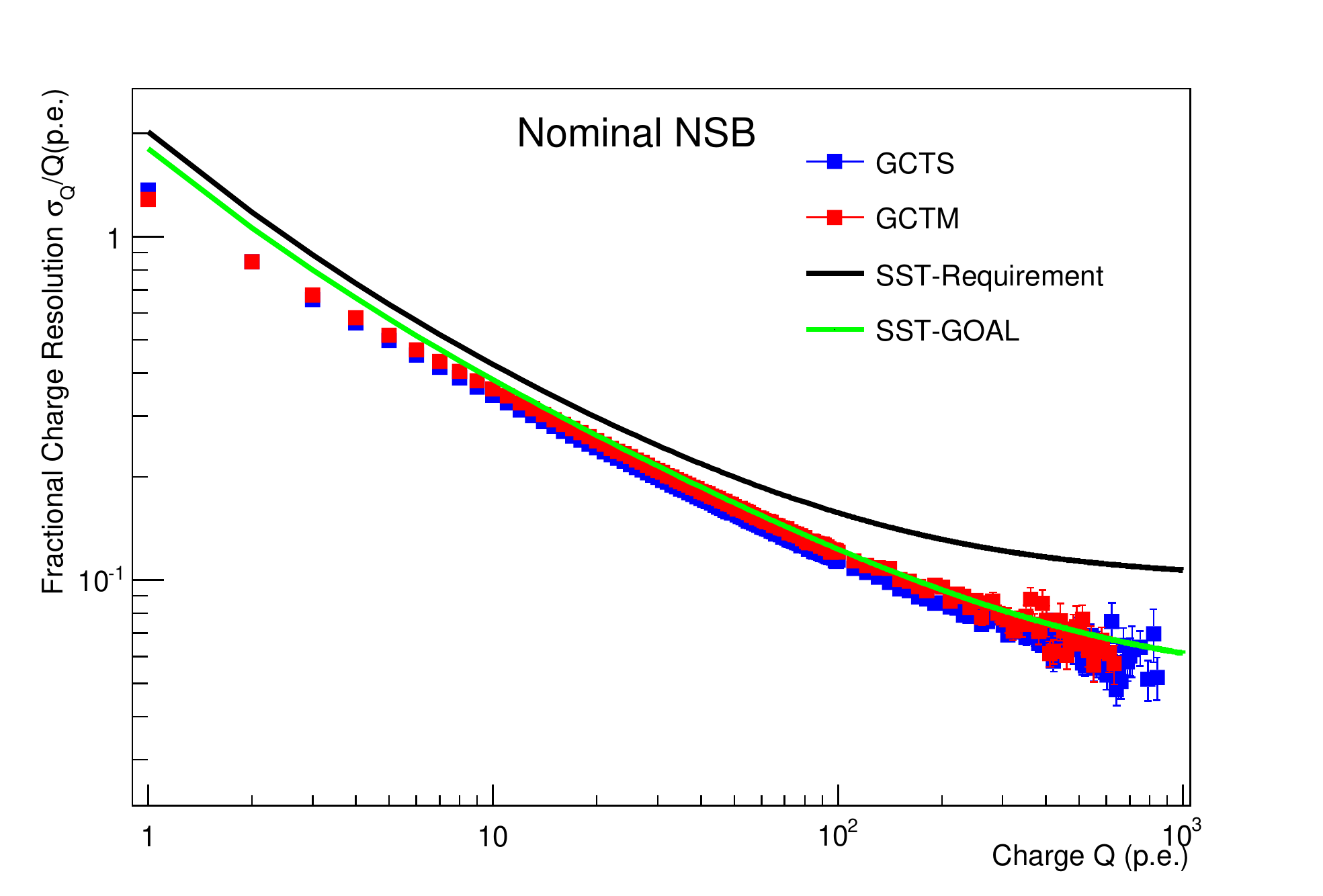} & \includegraphics[width=0.5\textwidth]{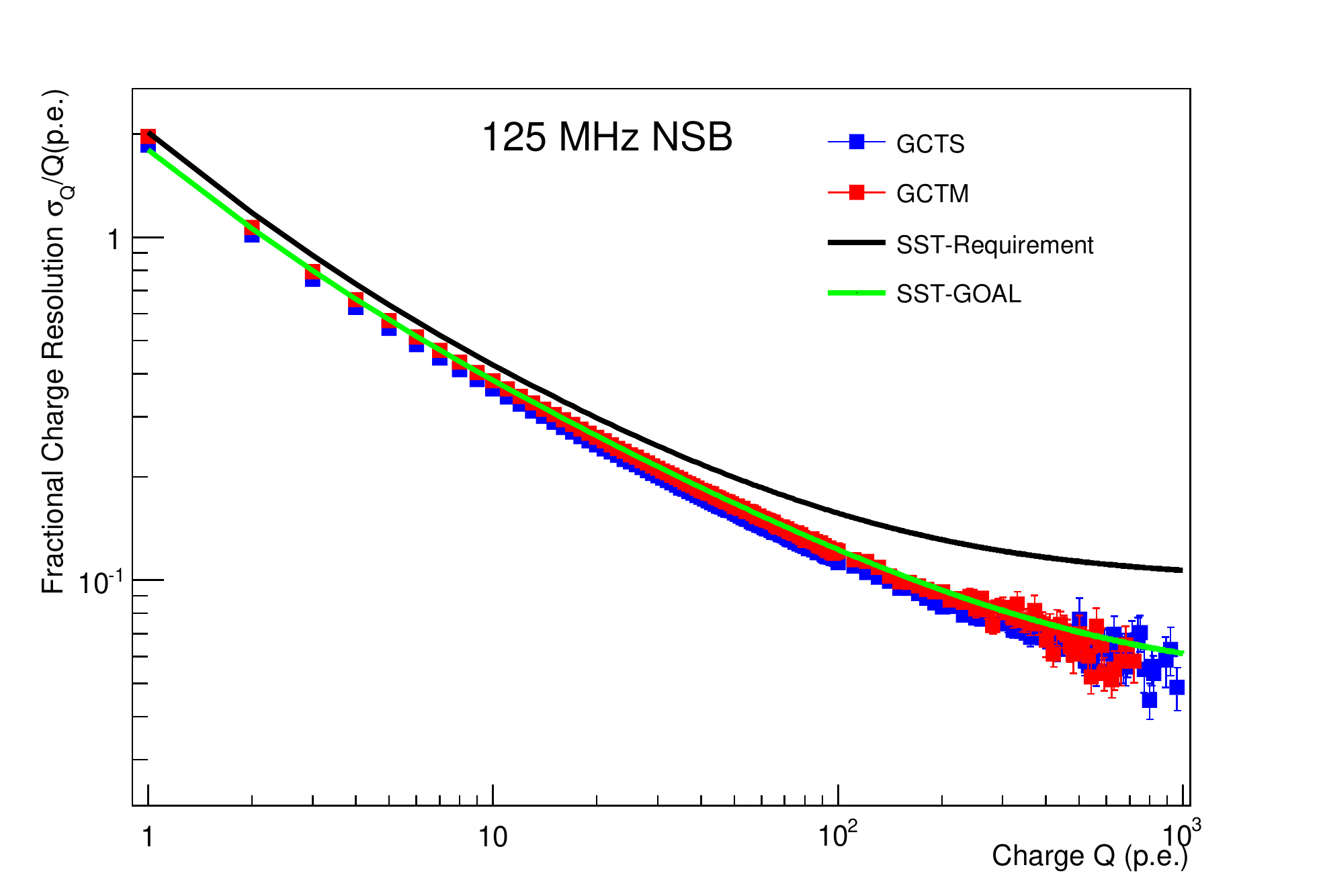} \\
%a) & b) \\
\end{tabular}
\caption{Fractional charge resolution of GCT-M (red)  and GCT-S (blue) as a function of true charge
in the case of nominal NSB scenario (left) and high NSB scenario (right).}
\label{fig:charge}
\end{figure}

Two different NSB scenarios have been considered. The nominal NSB scenario corresponds to standard dark sky conditions, while the high NSB condition corresponds to the maximum NSB considered in the CTA requirements. Figure \ref{fig:charge} shows the fractional charge resolution in these two scenarios. The black and the green lines represent the requirement and goal performance required by CTA for the SSTs. As expected, in the case of high NSB the performance is worse but it should still be acceptable to perform observations. Note that in these simulations we have not taken into account the saturation of the signal, which should significantly affect the charge resolution above 500 p.e.. However preliminary studies have demonstrated that the charge can be recovered using the FWHM of the waveform of the saturated pulses to estimate the true charge.

\section{Trigger Efficiency}
\label{sec:trigeff}

An important indicator of performance is how well the camera can trigger as a function of the mean integrated Cherenkov light intensity. The trigger logic of the camera is based on a combination of 4 neighbouring pixels being required to exceed the trigger threshold. One of the predefined performance requirements for our camera is that the trigger probability must be at least 50\% for an integrated camera charge of 100~p.e., with a goal of reaching below 80~p.e. 

\begin{figure}
\centerline{
\includegraphics[width=1.2\textwidth]{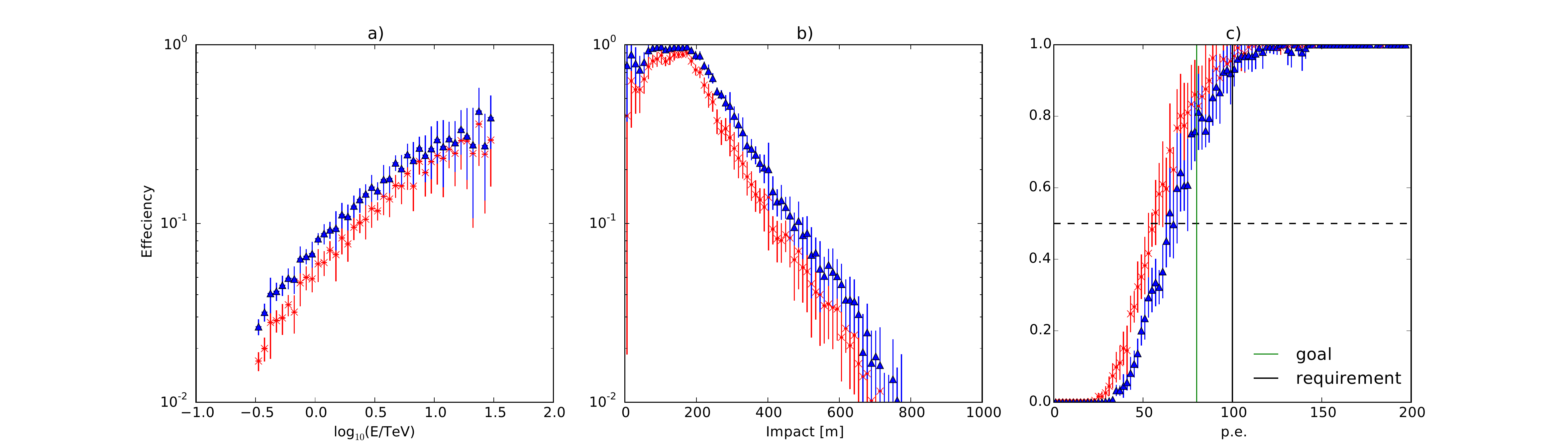}
}
\caption{Trigger efficiency of gamma rays for the two telescopes, GCT-M (red $\times$) and GCT-S (blue $\triangle$), as a function of a) Energy, b) Impact distance and c) photo-electrons.}
\label{fig:trigeff}
\end{figure}

In order to investigate this, we simulated $2\times10^{5}$ gamma-ray showers in the energy range 0.315 to 600~TeV out to a distance of 1200~m from a single GCT telescope. We then found the ratio of triggered showers over all showers for given charges. In Figure \ref{fig:trigeff}.c we can see that for both cameras the goal is exceeded, with GCT-M and GCT-S reaching a 50\% trigger efficiency at 57~p.e. and 67~p.e. respectively. In Figure \ref{fig:trigeff}.a and \ref{fig:trigeff}.b we also show the trigger efficiency as a function of both energy and impact distance; the latter confirms the idea that SSTs can be used at larger spacing than the Cherenkov light pool ($\sim$ 120~m) to improve the chance of triggering on these showers.

\section{Discussion and Outlook}

%The complete telescope structure has been mounted in Meudon. Aluminium petals of reduced size have been fabricated 
%for the M1. Their optical quality is currently being evaluated. The M2, equally made from aluminium in six pieces that will 
%be bolted together, is currently under construction. The MaPM based camera is currently undergoing commissioning in Leicester and has become the first CTA camera to see first light under lab conditions. Once mounted on the prototype later this year commissioning of the second, SiPM based prototype, will begin. With the knowledge gained from the prototypes, we have begun more detailed simulations in order to understand the final performance of the GCT telescope. 

The construction of the GCT-S prototype structure and camera is advancing rapidly. The mechanical assembly has been mounted at the Observatoire de Paris (Meudon, France) and the prototype aluminium M1 petals of reduced size have been fabricated. Their optical quality is currently being evaluated. The M2, equally made from aluminium in six pieces that will be bolted together, is currently under construction. The GCT-M camera is undergoing commissioning in Leicester and has become the first CTA camera to see first light under lab conditions. While the mirrors and first camera are placed on the structure later this year, commissioning of the second GCT-S camera should be underway. 

 With the knowledge gained from the prototypes, we have begun more detailed simulations in order to understand the performance of the final GCT telescope. We have considered the effects of shadowing, seeing an overall reduction in collection area of about 1~m$^{2}$ for on axis events. For the ideal PSF of the current design, we see that the 80\% containment radius falls below our pixel size for the entire field of view. Finally, both the charge resolution and trigger efficiency of the two cameras GCT-M and GCT-S meet the predefined requirements. As our knowledge of the system grows, we expect to achieve a better understanding of the performance of GCT as part of CTA as a whole. 

%The construction of the prototype telescope structure for GCT and the prototype camera is advancing rapidly, The complete telescope structure has been mounted in Meudon. Aluminium petals of reduced size have been fabricated for the M1 and their optical....

\vspace*{0.5cm}
\footnotesize{{\bf Acknowledgment:}{ Part of this work has been funded under the Convention 10022639 between the R\'{e}gion d'Ile-de-France and the Observatoire de Paris. We gratefully acknowledge the Ile-de-France, the CNRS (INSU and IN2P3), the CEA and the Observatoire de Paris for financial and technical support. We also acknowledge support from the agencies and organizations listed in this page: http://www.cta-observatory.org/?q=node/22. We would also like to acknowledge the support of the UK Science and Technology Facilities Council (grant  ST/K501979/1) } and Durham University. Finally we would like to thank ASTRI for their help with preparing the configurations for the SiPM detector modules}

\end{document}